\documentclass[
reprint,
 amsmath,amssymb,
 aps,
]{revtex4-2}

\usepackage{placeins} 
\usepackage{url}
\usepackage[hyperindex,breaklinks]{hyperref}
\hypersetup{linkcolor=blue, citecolor=red, colorlinks=true}

\usepackage{capt-of}
\usepackage{graphicx}
\usepackage{dcolumn}
\usepackage{bm}

\begin{document}

\title{
Enhanced spreading in continuous-time quantum walks \\ using aperiodic temporal modulation of defects
}

\author{José J. Ximenes$^{1}$}
\thanks{jeffersonximenes@ufu.br}

\author{Marcelo A. Pires$^{2}$}
\thanks{piresma@cbpf.br}

\author{José M. Villas-Bôas$^{1}$}
\thanks{boas@ufu.br}

\affiliation{
$^{1}$Instituto de Física, Universidade Federal de Uberlândia, 38400-902 Uberlândia-MG, Brazil
\\ 
$^{2}$Centro Brasileiro de Pesquisas F\'{\i}sicas, Rio de Janeiro - RJ, 22290-180, Brazil
}

\begin{abstract}  
Parrondo's paradox, where the alternation of two losing strategies can produce a winning outcome, has recently been demonstrated in continuous-time quantum walks (CTQWs) through periodic defect modulation. We extend this phenomenon to aperiodic protocols. We show that deterministic, non-repetitive defect switching can enhance quantum spreading in CTQWs compared to the defect-free case. Furthermore, we establish that the degree of this enhancement is strongly influenced by the autocorrelation and persistence characteristics of the applied aperiodic sequence. Our findings indicate that aperiodic defect modulation reliably maintains Parrondo's effect and provides new ways to control wavepacket properties in CTQWs.  
\end{abstract}
 
\keywords{a,b,c}
              
\maketitle


\section{Introduction}\label{sec:intro}

\begin{figure}[t]
    \centering
    \includegraphics[width=0.45\textwidth]{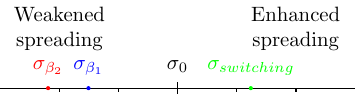}
\caption{
Illustration of the Parrondo's paradox in the realm of CTQWs that are subject to defects. To quantify the spreading of the CTQW, we compute the standard deviation, $\sigma$, of the wavepacket probability distribution.   Relative to the defect-free baseline ($\sigma_0$), spreading is classified as slow ($\sigma < \sigma_0$) or fast ($\sigma > \sigma_0$). The introduction of defects typically leads to a decrease in the spreading of the CTQW, exemplified by $ \sigma_{\beta_1}<\sigma_0$ (slow) and $ \sigma_{\beta_2}<\sigma_0$ (slow), representing the dynamics with a single transition defect. Surprisingly, proper temporal switching between $\beta_1$ and $\beta_2$ can lead to enhanced spreading, \textcolor{black}{$\sigma_{switching}>\sigma_0$} (fast). That is, the alternating of two unfavorable setups, can lead to favorable scenarios. This manifestation is termed the ``slow+slow$\rightarrow$ fast" effect. }
\label{fig:standardintro}
\end{figure}

The genesis of quantum walks can be traced back to the decade of the 1990s, with the introduction of two primary formulations: the discrete-time quantum walk (DTQW) and the continuous-time quantum walk (CTQW). The DTQW, formally introduced in 1993 by extending the classical random walk to the quantum realm~\cite{aharonov1993quantum}, proceeds in discrete time steps governed by the application of the coin and translation operators. Subsequently, in 1998, the  CTQW was also introduced~\cite{farhi1998firstCTQW}. The CTQW evolves over time under the influence of a Hamiltonian operator. The scientific interest in QWs is sustained by the various physical realizations~\cite{wang2013physical}, and by the ever-increasing range of applications~\cite{portugal2013quantum,kadian2021quantum,mulken2011continuous,khalilipour2025review,desdentado2025quantum}.

In parallel to the development of quantum walks, the field of game theory witnessed the emergence of intriguing phenomena such as the Parrondo's paradox, first formalized in 1999~\cite{HarmerAbbott1999}. In this counterintuitive phenomenon two losing games can surprisingly yield a gain when played in an alternating way. Parrondo's paradox has since been explored in various areas of science~\cite{abbott2010asymmetry,wen2024parrondo}. 

The intersection of these two aforementioned domains has been explored with several protocols for obtaining Parrondo-like effects in
QWs~\cite{flitney2004quantum,Kosik2007,chandrashekar2011parrondo,Rajendran2018EPL,rajendran2018implementing,MachidaGrunbaum2018,walczak2023noise,walczak2022parrondo,walczak2021parrondo,trautmann2022parrondo,lai2020parrondoCointoss,lai2020parrondo4sided,mielke2023quantum,pires2020parrondo,panda2022generating,jan2023territories,janexperimental2020,mittal2024,walczak2024parrondo,kadiri2024scouring,hosaka2024parrondo,laiparrondo2020Review}. But all of these earlier quantum manifestations of Parrondo's paradox have been formulated within the framework of the DTQW.  
Recently, in Ref.~\cite{ximenes2024parrondo},  Parrondo's paradox was introduced to the   domain of CTQWs. In this article, the authors  focused on periodic alternations of defects. Here we investigate, for the first time, how the application of aperiodically modulated defects can induce this paradoxical phenomenon in CTQWs.

As illustrated in Fig.~\ref{fig:standardintro}, the  Parrondo's paradox in CTQWs emerges when we detect~\cite{ximenes2024parrondo}:
    \begin{align}
        \sigma_{\beta_1} < \sigma_0   \\ 
        \sigma_{\beta_2} < \sigma_0 \\ 
        \sigma_{\text{switching}} > \sigma_0
    \end{align}

Beyond conventional periodic order, systems displaying nontrivial arrangements, such as aperiodic sequences~\cite{allouche2003automatic}, possess wide-ranging implications in several fields such as
condensed matter physics and chemistry~\cite{macia2023alloy} as well in non-equilibrium  phenomena~\cite{pires2022randomness,fernandes2023contact,barghathi2014contact}, stochastic game theory~\cite{luck2019parrondo,pires2024parrondo} and quantum walks~\cite{ribeiro2004aperiodic,romanelli2009fibonacci,di2015massless,andrade2018discrete,gullo2017dynamics,pires2020quantum,bose2024influence}.

The manuscript is organized as follows: in Sect.~\ref{sec:model} we describe our model designed to incorporate aperiodic defect alternation; In Sec.~\ref{sec:results} we present and discuss the results for our mathematically-designed physical system; and  in Sec.~\ref{sec:remaks} we offer final remarks on the broader significance of our findings. As we will show, the core contributions of this paper are threefold:
\begin{enumerate} 
  \item[(i)] We establish a novel application of aperiodic sequences within  quantum systems, specifically for defect modulation in CTQWs.
    \item[(ii)] We show that the Parrondo effect in CTQWs is robust and extends beyond periodic protocols to include aperiodic modulation of defects.
    \item[(iii)]  Our findings reveal that the structural properties of aperiodic sequences (e.g., autocorrelation, persistence) directly influence quantum transport, offering a new avenue for engineering spreading and delocalization in CTQWs.
\end{enumerate}

\section{Model} \label{sec:model}

\subsection{Continuous-time quantum walks}\label{sec:CTQW}

The system consists of a CTQW for a single particle on a one-dimensional lattice with two nonperiodic alternating transition defects. The model presented here extends the model developed in~\cite{ximenes2024parrondo}. Hence, we first define the Hamiltonian of defect-free QW, which propagates over sites ${ |j \rangle} $, as
\begin{equation}\label{H0}
    H_0 = \epsilon \sum_{j} |j \rangle \langle j| - \gamma \sum_{j} \left( |j + 1 \rangle \langle j| + |j - 1 \rangle \langle j| \right) ,
\end{equation}
with $\epsilon$ as the constant potential energy and $\gamma$ as the transition rate. For the purposes of this article, we set $\epsilon=0$, all the other variables are expressed in \textcolor{black}{units of the} $ \gamma $. As also done by Li and Wang~\cite{li2015single}, the term for additional defect transition for nearest neighbors of site $j=d$ is defined as
	\textcolor{black}{(i.e., times are reported in units of $\gamma^{-1}$ and energies in units of $\gamma$).}
\begin{equation}\label{Hb}
H_d = -  \left( |d \rangle \langle d+1| +|d + 1 \rangle \langle d| + |d - 1 \rangle \langle d|+ |d \rangle \langle d-1| \right) .
\end{equation}

{\color{black}
To alternate defect intensities in time we define the time-dependent function $f(t)$ driven by a binary control sequence $\{w_n\}$ with $w_n\in\{0,1\}$. We adopt the explicit representation
\begin{equation}
f(t)=w_{\lfloor t/\tau\rfloor}\,\beta_1 + \bigl(1-w_{\lfloor t/\tau\rfloor}\bigr)\,\beta_2,\qquad t\ge0,
\end{equation}
where $\lfloor\cdot\rfloor$ is the floor function, $\tau>0$ is the switching interval, and $\beta_1,\beta_2$ are the two defect intensities. Thus $f(t)$ is piecewise constant on intervals $[n\tau,(n+1)\tau)$: when $w_n=1$ the defect intensity is $\beta_1$, and when $w_n=0$ it is $\beta_2$ as illustrated in Fig.\ \ref{fig:switch}. The sequence index advances as $n=\lfloor t/\tau\rfloor$, so switching occurs at $t=n\tau$.
\begin{figure}[htb]
    \centering
    \includegraphics[width=1\columnwidth]{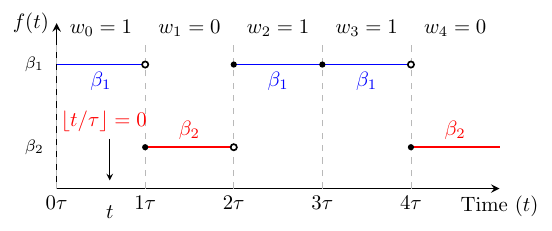}
    \caption{\textcolor{black}{Schematic of the switching protocol for the Binary sequence: $w_n = \{1, 0, 1, 1, 0\}$. The function $f(t)$ is piecewise constant on intervals $[n\tau,(n+1)\tau)$; at $t=n\tau$ the right-hand value is taken. Open circles mark excluded (left) values and filled circles mark included (right) values.}}
    \label{fig:switch}
\end{figure}
}

\color{black}
In contrast to Ref.~\cite{ximenes2024parrondo}, the switching function $f(t)$ is aperiodic. The time-dependent Hamiltonian is
\begin{equation}\label{Hp}
    H(t)=H_0+f(t)\,H_d.
\end{equation}
Following Ref.~\cite{ximenes2024parrondo}, we take $\beta_1=-2.5\gamma$ and $\beta_2=-3\gamma$ (other values produce the same qualitative phenomena), and set $d=0$ without loss of generality.

With $\hbar=1$, the state evolves according to
\begin{equation}\label{Psit}
    i\,\frac{\partial}{\partial t}\,|\Psi(t)\rangle=H(t)\,|\Psi(t)\rangle.
\end{equation}
The probability distribution is $P_j(t)=|\langle j|\Psi(t)\rangle|^2$ with $\sum_j P_j(t)=1$. Define the mean $\mu=\overline{j}=\sum_j j\,P_j(t)$ and, more generally, $\overline{g(j)}=\sum_j g(j)\,P_j(t)$. The standard deviation is
\begin{equation}\label{eq:std}
    \sigma=\sqrt{\overline{j^2}-\overline{j}^2}.
\end{equation}
	\textcolor{black}{This is equivalent to the variance definition $\sigma^2=\sum_j (j-\mu)^2 P_j$, since $\sum_j (j-\mu)^2 P_j=\overline{j^2}-\mu^2$ with $\mu=\overline{j}$.}

We also use the Shannon entropy
\begin{equation}\label{Shannon}
    S=-\sum_j P_j\,\log_{10} P_j,
\end{equation}
and the inverse participation ratio (we adopt the convention)
\begin{equation}\label{IPR}
    \mathrm{IPR}=\left(\sum_j P_j^2\right)^{-1}.
\end{equation}
For a localized wavepacket, $S=0$ and $\mathrm{IPR}=1$; for a uniform distribution over $N$ sites, $S=\log_{10}N$ and $\mathrm{IPR}=N$.

\color{black}

\subsection{Switching protocols}\label{subsec:switchingprotocols}

\begin{figure}[!htb]
 \centering
\includegraphics[width=1\linewidth]{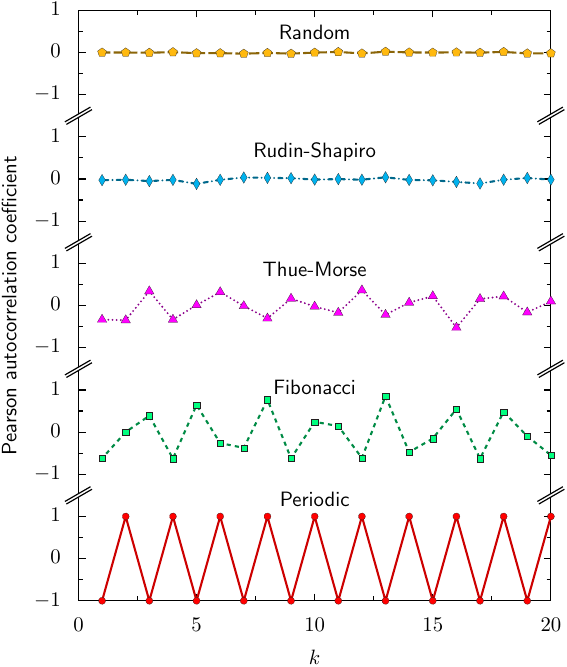}
\caption{Characterization of the binary sequences with the Pearson autocorrelation coefficient (AC)   for a given \textcolor{black}{lag $k>0$}. }
    \label{fig:aperiodic_seq_AC}
\end{figure}

\begin{figure}[!htb]
 \centering
 \includegraphics[width=1\linewidth]{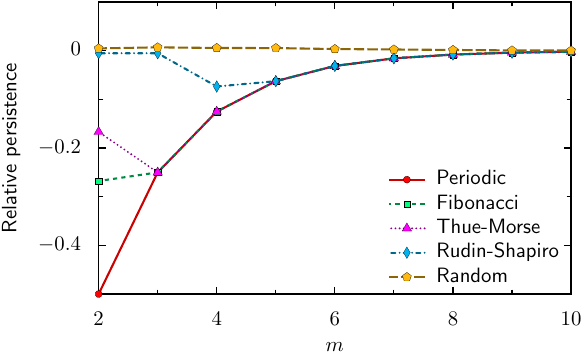} 
\caption{Characterization of the binary sequences with the   relative  persistence (RP) for a given  block size $m$. }
    \label{fig:aperiodic_seq_RP}
\end{figure}

In line with the standard literature on binary  sequences~\cite{steurer2007photonic,dal2012deterministic} we generate our aperiodic sequences with the substitution rules:  
\begin{enumerate}
    \item Fibonacci \textbf{(Fb)}: $w^{(0)} = 0$; $0 \mapsto 01$, $1 \mapsto 0$;
    \item Thue-Morse \textbf{(TM)}: $w^{(0)} = 0$; $0 \mapsto 01$, $1 \mapsto 10$; 
    \item Rudin-Shapiro \textbf{(RS)}: $w^{(0)} = A$; $A \mapsto AB$, $B \mapsto AC$, $C \mapsto DB$, $D \mapsto DC$; Binary conversion: $A, B \rightarrow 0$, $C, D \rightarrow 1$.
\end{enumerate}

\textcolor{black}{For further details see our appendix~\ref{sec:app_bin_seq}. } We adopt the abbreviations Pe and Rd to represent the periodic and random protocols, respectively.

\textcolor{black}{
We characterize binary sequences through their autocorrelation and  relative persistence. To provide a basis for comparison, we derive exact analytical expressions for these measures for canonical cases (e.g., periodic and random sequences) in Appendix~\ref{sec:app_exact_binseq}.
}

\color{black}
We first compute the Pearson autocorrelation coefficient for a   temporal lag $k$ using
\begin{align}
    AC(k) = \frac{\text{Cov}(w_t, w_{t+k})}{\sqrt{\text{Var}(w_t) \cdot \text{Var}(w_{t+k})}}
\end{align}
where $\text{Var}(\cdot)$ is the variance and $\text{Cov}(\cdot, \cdot)$ is the covariance.
\color{black}

Following Ref.~\cite{pires2024parrondo}, we   compute the   binary persistence of order $m$ given by
\begin{align}
    BP(m) = \frac{ N_{\text{ID}}(m) }{ N_{\text{T}}(m) } 
    \label{eq:bin_per}
\end{align}
where $N_{\text{ID}}(m)$ is the number of blocks where all elements are identical and $N_{\text{T}}(m)$ is the total number of blocks examined. That is, $BP(m)$ computes the probability of observing a subsequence of length $m$ consisting entirely of zeros or ones. For instance, $BP(3)$ corresponds to the proportion of subsequences $000$ or $111$ among all possible consecutive triplets in the sequence. 
 
As a baseline reference, we consider the random sequence. Then we compute the relative binary persistence
\begin{align}
    RP(m) = BP(m)  - 2^{1-m}
    \label{eq:rel_bin_per_v2}
\end{align}
This quantity captures the nontrivial deviation from randomness and has a clear interpretation: (a) $RP > 0 $ means a tendency to persistence of values, (b) $RP = 0$ means no persistence of patterns, (c) $RP < 0$ means an anti-persistent behavior, where the values tend to alternate frequently.

Figures~\ref{fig:aperiodic_seq_AC} and \ref{fig:aperiodic_seq_RP} illustrate a clear hierarchy in terms of overall autocorrelation and binary persistence as indicated by
\begin{align}
|AC_{Pe}| &> |AC_{Fb}| > |AC_{TM}| > |AC_{RS}| \approx |AC_{Rd}|  \\
|RP_{Pe}| &> |RP_{Fb}| > |RP_{TM}| > |RP_{RS}| > |RP_{Rd}| 
\end{align}
The negative values for $AC$ and $RP$ mean a propensity for a switching of values.

Note that autocorrelation alone is not a sufficient discriminator for these sequences, underscoring the importance of the complementary measure $RP$.

\begin{figure}[t]
    \centering
    \includegraphics[width=1\linewidth]{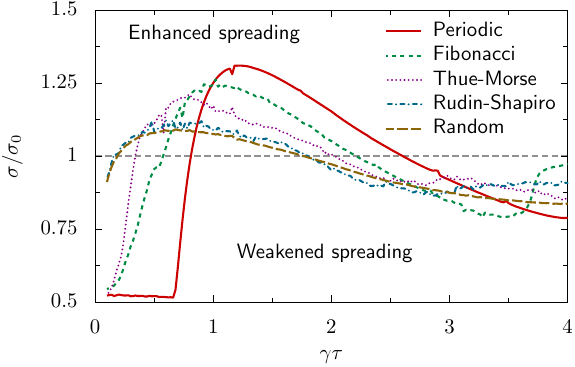}
\caption{Relative standard deviation $ \sigma/\sigma_0 $ as a function of $ \tau $. The point corresponds to dynamics evolution   for \textcolor{black}{$\gamma t = 2000$}. Each point on the random case curve represents the mean of 50 random sequences, therefore, the point number is smaller than that of the aperiodic deterministic curves. The enhanced spreading zone is separated by a horizontal dashed line from reduced spreading.  Note that there is an optimal switching interval,  $\tau^{\text{min}} < \tau < \tau^{\text{max}}$,  where defect modulation leads to the Parrondo's paradox  in the form "slow+slow$\rightarrow$ fast". Also note that the aperiodic protocol curves are not just a mere vertical shift of the periodic protocol curve.   }
    \label{fig:sigmavstau}
\end{figure}

\section{Results and discussion} \label{sec:results}

In this section, we present our finding regarding our model that bridges Parrondo's paradox, CTQWs and aperiodic design of defect modulation.

Figure \ref{fig:sigmavstau} shows that the Parrondo's effect in the CTQW is not restricted to a periodic protocol, but remains robust in the aperiodic setting. This figure reveals regions where the presence of these alternating defects leads to 
(i) an enhancement of quantum spreading ($\sigma > \sigma_0$) and 
(ii) a weakening of spreading ($\sigma < \sigma_0$). 
In all cases the occurrence of this phenomenon is nonmonotonic with $\tau$, size of the switching interval. That is, there exists a range   $\tau^{\text{min}} < \tau < \tau^{\text{max}} $ for detection of the paradoxical behavior. Outside this interval, when $\tau$ is either too short ($\tau < \tau^{\text{min}}$) or too long ($\tau > \tau^{\text{max}}$), the phenomenon disappears. This suggests that the system requires sufficient time in each defect state to manifest the behavior, but not so much that the switching becomes negligible. For  $\tau \in [ \tau^{\text{min}}, \tau^{\text{max}}  ] $ there is a maximum such that
\begin{align}
 \sigma_{Pe}^{max} > \sigma_{Fb}^{max} > \sigma_{TM}^{max} > \sigma_{RS}^{max} > \overline{\sigma}_{Rd}^{max}   
\end{align}
which is in accordance with the hierarchy of autocorrelation and persistence previously described.

In order to better understand the observed phenomena, let us take a look at the temporal properties of our proposed model.

\begin{figure}[t]
    \centering
    \includegraphics[width=1\linewidth]{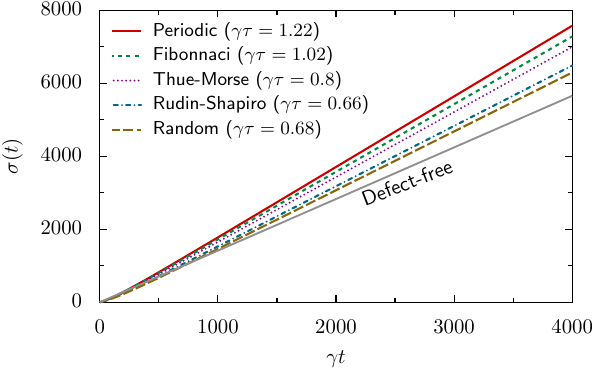}
\caption{Time evolution of the standard deviation for the defect-free case, along with periodic and aperiodic alternation between two defect values. The values of $\tau$ used here correspond to the maxima for each sequence in Fig. \ref{fig:sigmavstau}.}
    \label{fig:Standarvst}
\end{figure}

\begin{figure}[!htb]
    \centering
    \includegraphics[width=1\linewidth]{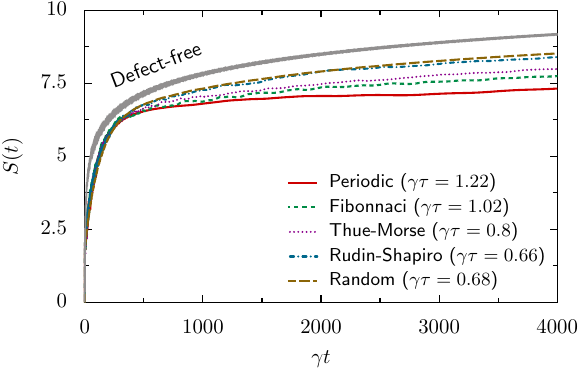}    
\caption{ Time evolution of the Shannon entropy for the evolutions in Fig.~\ref{fig:Standarvst}. At \textcolor{black}{\( \gamma t = 4000 \)}, higher spatial spreading (larger \(\sigma\)) correlates with lower entropy values, signifying a less uniform probability distribution despite the broader spatial extent.
}
\label{fig:Shannonvst}
\end{figure}

\begin{figure}[!htb]
\centering
\includegraphics[width=1\linewidth]{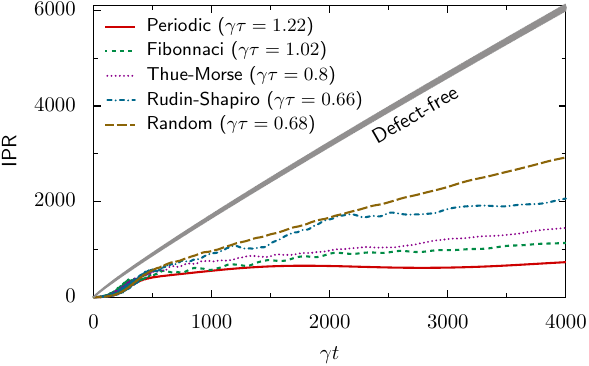}
\caption{Time evolution of the inverse participation ratio (IPR) for the evolutions in Fig.~\ref{fig:Standarvst}. At \textcolor{black}{\( \gamma t = 4000 \)}, higher spatial spreading (larger \(\sigma\)) correlates with lower IPR values, indicating increased probability concentration at fewer sites despite broader spatial distribution.
}
    \label{fig:IPRvst}
\end{figure}

Figure~\ref{fig:Standarvst} illustrates the time evolution of the standard deviation of the wavepacket's position. The results clearly indicate that in the long-time
\begin{align}
 \sigma_{Pe} > \sigma_{Fb} > \sigma_{TM} > \sigma_{RS} > \overline{\sigma}_{Rd}.
\end{align}
This  ordering is also consistent with the hierarchical values for the autocorrelation and persistence. 
While the transient behavior displays  different local slopes, we consistently observe the preservation of the ballistic-like regime in the long-run  for all protocols we analyzed.

Figures~\ref{fig:Shannonvst} and \ref{fig:IPRvst} illustrate the complementary aspects of delocalization through the Shannon entropy ($S$) and inverse participation ratio ($IPR$). These measures reveal an important trade-off: typically protocols yielding higher spatial spreading (larger $\sigma$) correlate with lower entropy values (indicating less uniform probability distributions) and lower IPR values (signifying increased probability concentration at fewer sites). 

This inverse relationship shows that enhanced spatial spreading under defect modulation comes with structured probability distributions rather than uniform delocalization. Despite their different mathematical formulations ($S$ has logarithmic dependence while $IPR$ has quadratic dependence), both measures consistently show the same hierarchy in the long-time limit:
\begin{align}
 {S}_{Pe} &< {S}_{Fb} < {S}_{TM} < {S}_{RS} < \overline{S}_{Rd}   \\
\mathrm{IPR}_{Pe} &< \mathrm{IPR}_{Fb} < \mathrm{IPR}_{TM} < \mathrm{IPR}_{RS} < \overline{\mathrm{IPR}}_{Rd}.
\end{align} 

That is, aperiodic defect modulation produces intermediate levels of delocalization situated between the corresponding results for the defect-free and periodic arrangements, with both entropy and IPR capturing the same hierarchical ordering as the spatial spreading.

In a unified perspective, the results of spreading (quantified by $\sigma$) and delocalization (quantified by $S$ and $\mathrm{IPR}$) indicate that the modulation of aperiodic defects maintains Parrondo's paradox and yields tunable spreading and delocalization, placed between defect-free and periodic arrangements. This intermediate behavior can be explained through the intermediate arrangement properties of the aperiodic sequences (quantified by the autocorrelation and binary persistence). In Appendix~\ref{sec:app_extra} we provide additional results that complement our findings.

\section{Final remarks}\label{sec:remaks} 

Different from previous related work that analyzed defects in QWs~\cite{childs2002example,zhang2014one,keating2007localization,agliari2010quantum,izaac2013continuous,benedetti2019continuous,li2013position,li2015analytical,da2021localization,kiumi2021eigenvalues}, we uncover that proper aperiodic temporal modulation of defects can induce a Parrondo phenomenon in continuous-time quantum walks (CTQWs), manifesting as a ``slow + slow $\rightarrow$ fast'' effect. In this way, we go beyond the recent work ~\cite{ximenes2024parrondo} by showing that this counterintuitive effect is not restricted to periodic switching protocols, but is more general and can appear in non-repetitive and structured temporal modulations of defects.

Our findings reveal that the enhancement of quantum spreading is deeply connected with the autocorrelation and persistence properties of the underlying aperiodic protocol. These results align with previous literature showing that the nontrivial properties of the aperiodic sequences can lead to interesting phenomena such as 
enhancement of the capital gain in alternating classical games~\cite{pires2024parrondo}, novel routes to superdiffusion in DTQWs~\cite{pires2020quantum}, distinctive electronic~\cite{jagannathan2021fibonacci} and thermal~\cite{wu2025mathematically} properties.

From an implementation perspective, our deterministic protocols offer significant advantages: They avoid the statistical sampling requirements of random protocols while maintaining precise control over wavepacket properties. This makes them particularly suitable for experimental realization in quantum platforms.

In a recent review~\cite{wu2025mathematically} it was highlighted how mathematical sequences can be leveraged to engineer materials with tailored thermal features. 
Building on the concept of mathematically inspired design in physical systems, we present a novel application of binary aperiodic sequences 
(Fibonacci, Thue-Morse, and Rudin-Shapiro) 
to control transport in CTQWs. 

In future research endeavors, we will explore the development of protocols capable of controlling the scaling dynamics from slower-than-ballistic to faster-than-ballistic regimes. This has been achieved for DTQWs~\cite{pires2019multiple}, but remains a challenge for CTQWs.

\section*{Acknowledgments}
Two of us (J. J. X and J. M. V. B.) would like to thank FAPEMIG for the financial support.

\bibliography{main.bib}

\appendix

\color{black}

\section{Details on the binary sequences}\label{sec:app_bin_seq}

We provide detailed information about the generation of aperiodic binary sequences used throughout this work. The sequences are constructed using substitution rules following established methods in the literature~\cite{steurer2007photonic,dal2012deterministic}.
    
\subsection*{Fibonacci Sequence}

The Fibonacci sequence is generated using the following substitution rule:
\begin{itemize}
    \item $0 \mapsto 01$, $1 \mapsto 0$
\end{itemize}

The first generations of the Fibonacci sequence are:
\begin{align*}
w^{(0)} &= 0 \\
w^{(1)} &= 01 \\
w^{(2)} &= 010 \\
w^{(3)} &= 01001 \\
\ldots
\end{align*}
This sequence exhibits properties related to the golden ratio and possesses long-range order without periodicity.

\subsection*{Thue-Morse Sequence}

The Thue-Morse sequence is generated using the following substitution rule:
\begin{itemize}
    \item $0 \mapsto 01$, $1 \mapsto 10$
\end{itemize}

The first generations of the Thue-Morse sequence are:
\begin{align*}
w^{(0)} &= 0 \\
w^{(1)} &= 01 \\
w^{(2)} &= 0110 \\
w^{(3)} &= 01101001 \\
\ldots
\end{align*}
This sequence exhibits remarkable properties such as long-range correlations without periodicity and a balance between the number of 0s and 1s.

\subsection*{Rudin-Shapiro Sequence}

The Rudin-Shapiro sequence requires a four-letter alphabet with the following substitution rule:
\begin{itemize}
    \item $A \mapsto AB$, $B \mapsto AC$, $C \mapsto DB$, $D \mapsto DC$
    \item  $A, B \rightarrow 0$; $C, D \rightarrow 1$
\end{itemize}

The first symbolic generations are:
\begin{align*}
w^{(0)} &= A \\
w^{(1)} &= AB \\
w^{(2)} &= ABAC \\
w^{(3)} &= ABACABDB \\
\ldots
\end{align*}

Applying the binary conversion rules to the Rudin-Shapiro symbolic generations yields the binary sequence:
\begin{align*}
w^{(0)} &= 0 \\
w^{(1)} &= 00 \\
w^{(2)} &= 0001 \\
w^{(3)} &= 00010010 \\
\ldots
\end{align*}
A well-known property of the Rudin-Shapiro sequence is its lack of autocorrelation.

\color{black}

\section{Exact results for canonical sequences}
\label{sec:app_exact_binseq}

We provide closed-form results for the binary persistence and the autocorrelation for the periodic and the random sequences.

\subsection{Periodic sequence}

\color{black}
\subsubsection{Autocorrelation}

 We analyze the autocorrelation of a periodic binary sequence, $ \{0, 1, 0, 1, \ldots\}$. This sequence exhibits a mean of $E[w_t] = 1/2$ and a variance of $\text{Var}(w_t) = 1/4$. Given its stationary nature, the variance remains constant, i.e., $\text{Var}(w_{t+k}) = \text{Var}(w_t) = 1/4$.

The covariance function, $\text{Cov}(w_t, w_{t+k})$, is conditional on the temporal lag $k$. For even $k$, the elements are identical ($w_{t+k} = w_t$), and the covariance equals $\text{Cov}(w_t, w_{t+k}) = E[(w_t - 1/2)^2] = \text{Var}(w_t) = 1/4$. For odd $k$, the elements are inverted ($w_{t+k} = 1 - w_t$), and the covariance is $\text{Cov}(w_t, w_{t+k}) = E[(w_t - 1/2)(1/2 - w_t)] = -E[(w_t-1/2)^2] = -\text{Var}(w_t) = -1/4$.

Thus, the Pearson autocorrelation is
\begin{align}
AC(k) = \begin{cases}
1 & \text{if } k \text{ is even} \\
-1 & \text{if } k \text{ is odd}
\end{cases}
\end{align}

\color{black}

\subsubsection{Binary persistence}

 For $m = 1$, all blocks of length $1$ consist of either $0$ or $1$, so they are trivially identical. Then, $N_{\text{ID}}(1) = N_{\text{T}}(1) = \text{length of the sequence}$. Thus, $BP(1) = 1$.

For any $m > 1$, any subsequence of length $m$ will contain both $0$s and $1$s because the sequence alternates.  Example for $m=3$: possible blocks are $010$, $101$, which are not identical. Similarly, for larger $m$, the pattern continues to alternate. Thus, $N_{\text{ID}}(m) = 0$ for all $m \geq 2$. Therefore, $BP(m) =  0$.
  
In summary, for the periodic sequence we obtain
\begin{align}
    BP(m) = \begin{cases}
        1 & \text{if } m = 1, \\
        0 & \text{if } m \geq 2.
    \end{cases}
\end{align}

\subsection{Random sequence}

\color{black}
\subsubsection{Autocorrelation}


Since the values in a random sequence are independent, the covariance for any non-zero lag is zero. 
Then, the  autocorrelation coefficient for $k > 0$ is $AC(k) = 0$.

\color{black}

\subsubsection{Binary persistence}

For a binary random sequence with equal probabilities $p(0) = 0.5$ and $p(1) = 0.5$, we obtain:

\begin{enumerate}
    \item Probability of an all-0 block of length $m$:
    \begin{equation}
        P(0_m) = \prod_{i=1}^m p(0) = (0.5)^m
    \end{equation}
    
    \item Probability of an all-1 block of length $m$:
    \begin{equation}
        P(1_m) = \prod_{i=1}^m p(1) = (0.5)^m
    \end{equation}
    
    \item Probability of either an all-0 or all-1 block (identical block):
    \begin{equation}
        P_{\text{ID}}(m) = P(0_m) + P(1_m) = 2 \times (0.5)^m = (0.5)^{m-1}
    \end{equation}
     
\end{enumerate} 

In summary, for a sufficiently long sequence, the expected value of $BP(m)$ for a binary random sequence is:
\begin{align}
    BP(m) =2^{1-m}  
\end{align}

\section{Extra results  }\label{sec:app_extra}
\begin{figure}[!htb]
    \centering
    \includegraphics[width=0.49\textwidth]{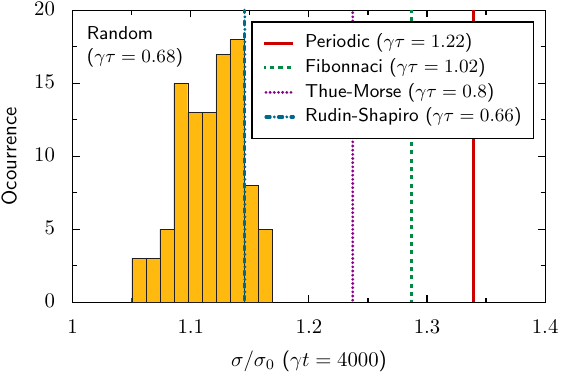}
\caption{Histogram of $\sigma/\sigma_0$ for 100 random sequences. The peak of the distribution (the most frequent $\sigma/\sigma_0$ value) is lower than that of the Rudin–Shapiro sequence, although some individual random realizations surpass it. The $\sigma/\sigma_0$ values for the other sequences are indicated by vertical lines for comparison.}
    \label{fig:histogram}
\end{figure}

Figure~\ref{fig:histogram} shows how the standard deviation of several random sequences is distributed around the average for the interval switching that produces the maximum spreading at time $\gamma t=4000$.

For this switching interval (which corresponds to the optimal $\tau$ for the random protocol, see Figure~\ref{fig:sigmavstau}), we observe that all $100$ random realizations exhibit enhanced spreading ($\sigma/\sigma_0 > 1$). The mean $\sigma/\sigma_0$ for the random sequences, as well as the values for the other sequences (indicated by vertical lines for their specific values of $\tau$), generally follows the hierarchy of autocorrelation and persistence. Although some individual random sequences can surpass the Rudin–Shapiro case, such events are rare.

\end{document}